\documentclass[draftcls, onecolumn,11pt]{IEEEtran}
\IEEEoverridecommandlockouts

\usepackage[english]{babel}
\usepackage{graphicx}
\usepackage[hypcap]{caption}
\usepackage{float}
\usepackage{multicol}
\usepackage{multirow}

\usepackage[usenames, dvipsnames]{color} 
\definecolor{mygray}{gray}{0.8}
	
\usepackage{float}
\usepackage{amssymb}
\usepackage{amsmath}
\usepackage{amsthm}
\usepackage{mathtools}
\usepackage{algorithm}
\usepackage[noend]{algpseudocode}

\usepackage[english]{babel}

\usepackage{subcaption}

\theoremstyle{remark}
\newtheorem{theorem}{Theorem}

\newtheorem{corollary}[theorem]{Corollary}
\newtheorem{definition}{Definition}
\newtheorem{lemma}{Lemma}

\newcommand{\mc}{\mathcal}

\ifCLASSINFOpdf
\else
\fi

\begin{document}
\title{
~\\Guesswork Subject to a Total Entropy Budget
}

\author{Arman Rezaee, Ahmad Beirami, Ali Makhdoumi, Muriel M\'edard, and Ken Duffy

\thanks{A. Rezaee, A. Beirami, A. Makhdoumi, and M. M\'edard are with the Research Laboratory of Electronics, Massachusetts Institute of Technology, Cambridge, MA, USA. Emails: \{armanr, beirami, makhdoum, medard\}@mit.edu.}

\thanks{K. Duffy is with the National University of Ireland Maynooth, Ireland. Email: ken.duffy@nuim.ie.}
}

\maketitle

\begin{abstract}
We consider an abstraction of computational security in password protected systems where a user draws a secret string of given length with i.i.d. characters from a finite alphabet, and an adversary would like to identify the secret string by querying, or guessing, the identity of the string. The concept of a ``total entropy budget'' on the chosen word by the user is natural, otherwise the chosen password would have arbitrary length and complexity. One intuitively expects that a password chosen from the uniform distribution is more secure. This is not the case, however, if we are considering only the average guesswork of the adversary when the user is subject to a total entropy budget. The optimality of the uniform distribution for the user's secret string holds when we have also a budget on the guessing adversary. We suppose that the user is subject to a ``total entropy budget'' for choosing the secret string, whereas the computational capability of the adversary is determined by his ``total guesswork budget.'' We study the regime where the adversary's chances are exponentially small in guessing the secret string chosen subject to a total entropy budget. We introduce a certain notion of uniformity and show that a more uniform source will provide better protection against the adversary in terms of his chances of success in guessing the secret string.
In contrast, the average number of queries that it takes the adversary to identify the secret string is smaller for the more uniform secret string subject to the same total entropy budget.
\end{abstract}

\IEEEpeerreviewmaketitle

\section{Introduction} \label{sec:Intro}
We consider the problem of identifying the realization of a discrete random variable $X$ by repeatedly asking questions of the form: ``Is x the identity of X?". This problem has been extensively studied by cryptanalysts who try to identify a secret key by exhaustively trying out all possible keys, where it is usually assumed that the secret key is drawn uniformly at random. We consider an $n$-tuple $X^n := X_1,\dots, X_n$ drawn from an i.i.d. source, $\mu_\theta(\cdot)$ on a finite alphabet $\mathcal{X},$ where $\theta$ represents the corresponding categorical distribution, which is not necessarily uniform. 
We measure security against a brute-force attacker who knows the source statistics completely, and who would query all the secret strings one by one until he is successful. 

Denoting the number of guesses by $G_\theta^n(X^n)$, the optimal strategy of the attacker that minimizes the expected number of queries $\mathop{\mathbb{E}}[G_\theta^n(X^n)]$ is to guess the possible realizations of $X^n$ in order of decreasing probability under $\mu_\theta^n(\cdot)$. Massey~\cite{massey1994guessing} proved that the Shannon entropy of $X^n$, $H(X^n)$, is a lower bound on the rate of growth of the expected guesswork, yet there is no upper bound on $\mathop{\mathbb{E}}[G_\theta^n(X^n)]$ in terms of $H(X^n)$. Ar{\i}kan~\cite{arikan1996inequality} proved that when we consider a string of growing length whose characters are drawn i.i.d, the positive moments of guesswork associated with the optimal strategy grow exponentially, and the exponents  are related to the R\'enyi entropies of the single letter distribution:\footnote{In this paper, $\log(\cdot)$ denotes the natural logarithm.}
\begin{eqnarray}
\lim_{n\to \infty}  \frac{1}{n} \log {\mathbb{E}}_{\theta}\left[ \left( G_{\theta}^n(X^n)\right)^\rho\right] = H_{1/(1+\rho)} \left(X\right),
\label{eq:metric1}
\end{eqnarray}
where the R\'enyi entropy of order $\rho$ is 
\begin{eqnarray}
H_{\rho} (X) = \frac{1}{1-\rho} \log\left( \sum_{x \in \mathcal{X}} P(X = x)^\rho \right).
\end{eqnarray}
Note that $\lim_{\rho \to 0} H_\rho(X) = H(X)$ recovers the Shannon entropy. We also use the notations $H_\rho(\theta)$ and $H_\rho(X)$ interchangeably to refer to the R\'enyi entropy of a string drawn from a source with parameter vector $\theta$. Although these connections have been extended to more general stochastic processes~\cite{sullivan-markov,sullivan-stationary}, in this paper, we focus on i.i.d. processes for the sake of clarity of presentation.

Christiansen and Duffy~\cite{christiansen2013guesswork} showed that the sequence $\{n^{-1} \log G_\theta^n(X^n)\}$ satisfies a Large Deviations Principle (LDP) and characterized its rate function, $\Lambda^{*}_\theta$. 
Beirami \textit{et al.}~\cite{beirami_allerton2015, Beirami-IT-Guesswork} showed that $\Lambda_\theta^*$ can be expressed as a parametric function of the value of a ``tilt'' in a family of tilted distributions.

We remark that when the metric of difficulty is the growth rate
in the expected number of guesses as a function of string length,
the challenge for the adversary remains the same even if the adversary
does not know the source statistics~\cite{sundaresan-universal, ISIT15_guesswork}.

In this paper, we first show a counter intuitive result that the average guesswork increases when the source becomes ``less uniform'' if the user is subject  to a total entropy budget on the secret string. Next, we introduce a natural notion of total guesswork budget on the attacker and show that the probability of success of an  adversary subject to a total guesswork budget increases when the source becomes ``less uniform,'' which is consistent with our intuition of choosing uniform passwords. We will formalize these notions in the rest of this paper.

\section{Problem Setup}
Given a finite alphabet $\mc{X}$,
a memoryless (i.i.d) source on $\mc{X}$ is defined by the set of probabilities $\theta_i = P[X = x_i]$ for all $i \in [|\mc{X}|]$, where $[n]:= \{1, \ldots, n\}$ and $\sum_{i \in [|\mc{X}|]} \theta_i =1$.  Hence, $\theta$ is an element of the $(|\mathcal{X}|-1)$-dimensional probability simplex. We define $\Theta_{|\mc{X}|}$ as the open set of all probability vectors $\theta$ such that $\theta_i > 0$ for all $i \in \{1,\dots,|\mathcal{X}|\}$, which also excludes the uniform source $u_{|\mc{X}|} = (1/|\mc{X}|, \ldots, 1/|\mc{X}|).$

The tilt operation plays a central role in the analysis, and is the basis for many of our derivations:
\begin{definition}[tilted $\theta$ of order $\alpha$~\cite{beirami_allerton2015}]
For any $\alpha \in\mathbb{R}$, define $\tau(\theta, {\alpha)}$ as the ``tilted $\theta$ of order $\alpha$'', where 
$\tau(\theta,\alpha) = (\tau_1(\theta,\alpha), \ldots, \tau_{|\mc{X}|} (\theta,\alpha))$, where $\tau_i: \Theta_{|\mc{X}|} \times \mathbb{R} \to \Theta_{|\mc{X}|}$ for all $i \in [|\mc{X}|]$ is given by
\begin{equation}
\tau_i(\theta,\alpha) := \frac{\theta_i^\alpha}{\sum_{\i=1}^{|\mc{X}|} \theta_i^\alpha}.
\end{equation}
\end{definition}

\begin{definition}[tilted family of $\theta$]
Let $\Gamma_\theta^+ \in \Theta_{|\mc{X}|}$ denote the ``tilted family of $\theta$'' and be given by
\begin{equation}
\Gamma_\theta^+:= \{\tau(\theta,\alpha) :{\alpha \in \mathbb{R}_{>0}}\}.
\end{equation}
\end{definition}
Observe that $\Gamma^{+}_\theta \in \Theta_{|\mc{X}|}$ is a continuum of stochastic vectors in the probability simplex.
Thus, the tilted family of a memoryless string-source with parameter vector $\theta$ is comprised of a set of memoryless string-sources whose parameter vectors belong to the tilted family of the vector $\theta$, i.e., $\Gamma^+_\theta$.

\begin{figure}[t]  
  \centering
  \includegraphics[width = 0.7\textwidth]{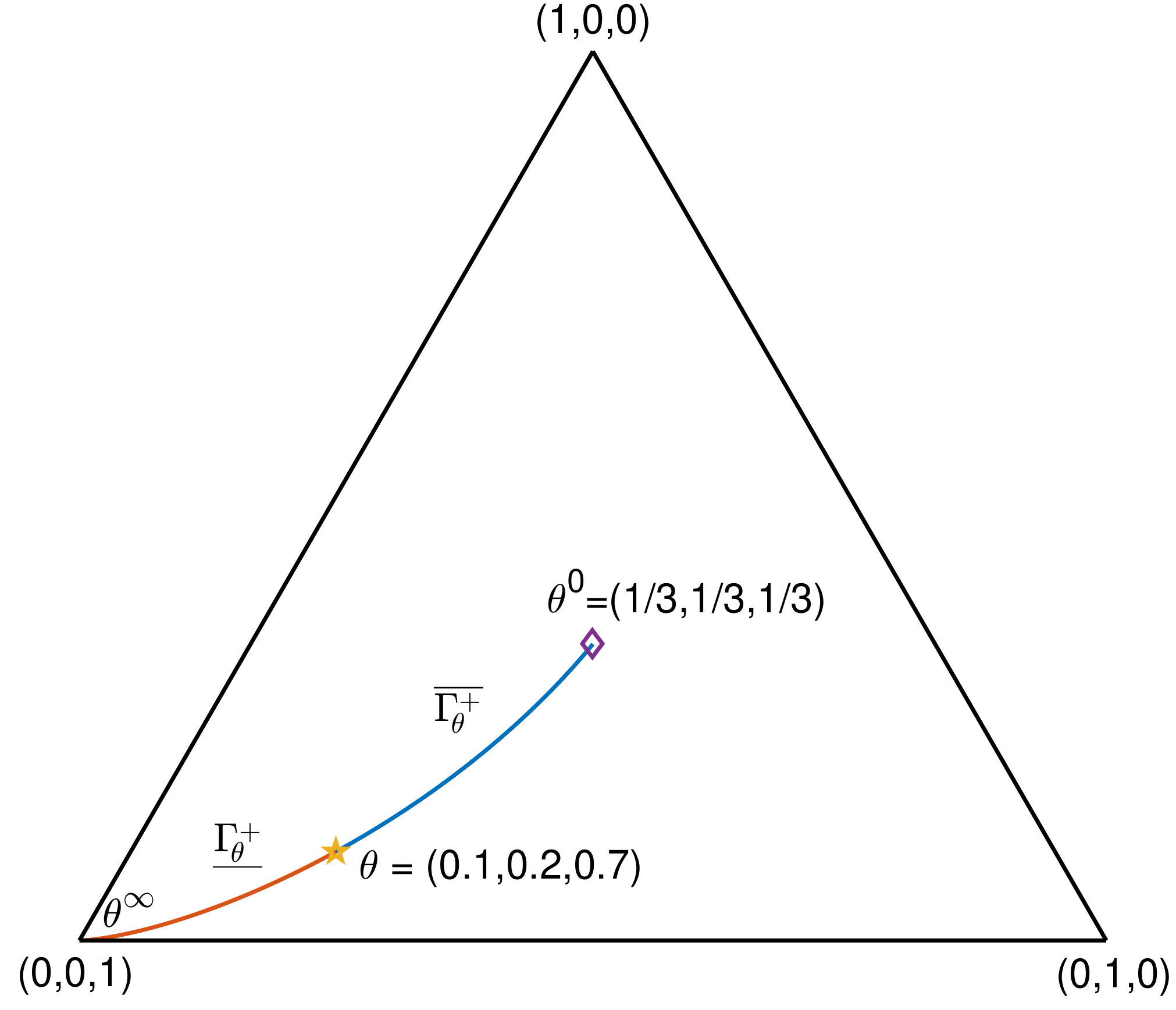}
  \caption{The probability simplex for a ternary alphabet. The figure represents the tilted family of $\theta=(0.1,0.2,0.7)$, as well as the high-entropy and low-entropy members of the family.}
\label{fig:Triangle}
\vspace{-.05in}
\end{figure}

\begin{definition}[high-entropy/low-entropy members of tilted family of $\theta$]
Let $\overline{\Gamma}_\theta^+$ and $\underline{\Gamma}_\theta^+$ denote the sets of high-entropy and low-entropy members of the tilted family of $\theta$, respectively, and be given by:
\begin{equation}
\overline{\Gamma}_\theta^+ = \left\{ \tau(\theta,\alpha) \right\}_{0\le\alpha < 1}, \hspace{1cm} \underline{\Gamma}_\theta^+ = \left\{ \tau(\theta,\alpha) \right\}_{\alpha > 1}.
\end{equation}
Hence,
$
\Gamma_\theta^+ = \overline{\Gamma}_\theta^+ \cup \underline{\Gamma}_\theta^+ \cup \theta . 
$
\end{definition}
Figure~\ref{fig:Triangle} depicts the probability simplex of all possible ternary parameter vectors, $|\mathcal{X}| = 3$. The yellow star represents the distribution $\theta = (0.1,0.2,0.7)$.  Note that the tilted family of $\theta$ is parametrized by $\alpha$. At $\alpha=0$, we get the uniform distribution $\tau(\theta,0) = u_3 = (1/3,1/3,1/3)$ and as $\alpha \to \infty$, we get to the degenerate case of $(0,0,1)$. The high-entropy and low-entropy members of the tilted family of $\theta$ are represented by blue and red, respectively. Note that all distributions in the high-entropy set, $\overline{\Gamma}_\theta^+$, have Shannon entropies higher than that of $\theta$ and are closer to the uniform distribution in the KL divergence sense~\cite{Beirami-IT-Guesswork}. Hence, the higher entropy members of the tilted family are ``more uniform'' than the lower entropy members of the tilted family.

\begin{definition}[entropy budget per source character]
Let $h \in (0,\log |\mc{X}|]$ denote the entropy budget per source character such that the user is required to choose a secret string from an i.i.d. process with parameter vector $\theta$ with $H(\theta) = h$.
\end{definition}
The concept of  a total entropy budget on the entire secret string is a natural one or the user would choose an arbitrarily complex secret string. We use the entropy budget per source character defined above to ensure that the user is subject to the same total entropy budget by adjusting the length of the secret string for a fair comparison between string sources that have different entropy rates.

\section{Positive Moments of Guesswork}
We first consider choosing strings with the same total (Shannon) entropy budget and measure security in terms of the positive moments of guesswork. If two sources have different entropy rates, we adjust the comparison by drawing a longer string from the lower entropy source. 
Formally, let us consider two sources with parameter vectors $\theta_1$ and $\theta_2$ on alphabet $\mc{X}$. 
Further, let $H(\theta_1)$ and $H(\theta_2)$ be the entropy rates of the two sources.
Let the entropy ratio be
\begin{equation}
\eta := \frac{H(\theta_2)}{H(\theta_1)}.
\label{eq:def-c}
\end{equation}
Without loss of generality, throughout this paper we assume that $H(\theta_2)<H(\theta_1)$, and hence $0<\eta<1$. The user is given the option to choose a secret string from either of the two sources. For a fair comparison, we assume that the entropy of the two strings is the same, $n_1 H(\theta_1) = n_2 H(\theta_2)$.
That is
\begin{align}
n_2& = \frac{1}{\eta} n_1.
\label{eq:ratio}
\end{align}
To compare the growth rates of the positive moments of guesswork, in light of~\eqref{eq:metric1}, we compare $H_{1/(1+\rho)}(\theta_1)$ and $\frac{1}{\eta} H_{1/(1+\rho)}(\theta_2)$. This will in turn impose the same total entropy budget on the strings drawn from the sources with parameter vectors $\theta_1$ and $\theta_2$.

For a parameter vector $\theta$, let an information random variable be defined as one that it takes the value $\log \frac{1}{\theta_i}$ with probability $\theta_i$ for all $i \in [|\mc{X}|]$. 
We need one more definition before we can state the result of this section:
\begin{definition}[skewentropy condition (SEC)]
A source with parameter vector $\theta \in \Theta_{|\mc{X}|}$ is said to satisfy the skewentropy condition (SEC) if
\begin{equation}
V(\theta)^2 + 2H(\theta) V(\theta) - H(\theta) S(\theta) >0,
\end{equation}
where $V(\theta)$ is the varentropy defined as the variance of an information random variable corresponding to $\theta$:
\begin{equation}
V(\theta) := \sum_{i \in [|\mc{X}|]} \theta_i \left(\log \frac{1}{\theta_i} - H(\theta) \right)^2.
\label{eq:varentropy}
\end{equation}
and $S(\theta)$ is the skewentropy, which is the skewness of an information random variable corresponding to $\theta$:
\begin{equation}
S(\theta) := \sum_{i \in [|\mc{X}|]} \theta_i \left(\log \frac{1}{\theta_i} - H(\theta) \right)^3.
\label{eq:skewness}
\end{equation}
\label{def:information-skewness}
\end{definition}
Note that varentropy has been studied extensively and naturally arises in the finite block length information theory~\cite{strassen, polyanskiy-finite-block}, and more recently in the study of polar codes~\cite{arikan-varentropy}. To the best of our knowledge, skewentropy has not been studied before, and we provide some properties of the SEC in Section~\ref{sec:skewness}. 

Equipped with this definition, we provide an ordering of the sources that belong to the same tilted family.

\begin{theorem}
Let $\theta_1 \in \Theta_{|\mc{X}|}$. For any $\theta_2 \in \underline{\Gamma}^+_{\theta_1}$,
\begin{equation}
H_{1/(1+\rho)}(\theta_1) < \frac{1}{\eta} H_{1/(1+\rho)}(\theta_2) \quad \forall \rho>0,
\end{equation}
if and only if $\theta_1$ satisfies the SEC in Definition~\ref{def:information-skewness}.
Note that  $\eta$ is the entropy ratio defined in~\eqref{eq:def-c}.
\label{thm:avg_guesswork}
\end{theorem}
The proof is provided in the appendix.
Theorem~\ref{thm:avg_guesswork} provides a natural ordering of sources that belong to the same tilted family. The ``less uniform'' low per-character entropy members of the tilted family take exponentially more number of queries, on the average, to breach compared to their more uniform higher per character entropy counterparts. 
\begin{corollary}
Let $u_{|\mc{X}|}$ denote the uniform source. Then for any $\theta \in \Theta_{|\mc{X}|}$, and any $\rho>0$,
$$
\log |\mc{X}| = H_{1/(1+\rho)}\left(u_{|\mc{X}|}\right) < \frac{1}{\eta} H_{1/(1+ \rho)}(\theta),
$$
where $\eta= H(\theta)/ \log |\mc{X}|.$
\label{thm:H-uniform}
\end{corollary}
Corollary~\ref{thm:H-uniform} suggests that, of all sources whose parameter vectors are in the (interior of the) probability simplex, the uniform source is the easiest to breach in terms of the positive moments of guesswork when the user is subject to a total entropy budget.
This is in contrast to our intuition that more uniformity provides better security.

\section{Probability of Success subject to
a Guesswork Budget}
In this section, we put forth a natural notion of total guesswork budget, leading to a security metric consistent with our intuition. Similar to the case of an entropy budget, we need to define guesswork budget per source character for our analysis.
\begin{definition}[guesswork budget per source character]
Let $ g \in (0,\log |\mc{X}|]$ denote the guesswork budget per source character, such that $e^{gn}$ is the total number of queries that the inquisitor can make in order to identify a secret string of length $n$. 
\end{definition}
Note that by this definition, the inquisitor is supposed to possess the resources for querying an exponentially growing number of strings (with the sequence length). In particular,  $g = \log|\mc{X}|$ corresponds to an adversary who is capable of querying all of the possible $|\mc{X}|^n$ outcomes of the source to successfully identify the secret string with probability $1$.

\begin{lemma}
If $g<H(\theta)$, then 
$$\lim_{n \to \infty} \mathbb{P}_\theta[G_\theta(X^n) \leq e^{gn}] = 0,$$
and if $g>H(\theta)$, then
$$\lim_{n \to \infty} \mathbb{P}_\theta[G_\theta(X^n) \leq e^{gn}] = 1.$$
\label{lem:shannon-entropy}
\end{lemma}
Recall that Ar{\i}kan~\cite{arikan1996inequality} showed that the growth rate of the moments of guesswork is governed by atypical sequences resulting in the appearance of the R\'enyi entropies in the expression. On the other hand, Lemma~\ref{lem:shannon-entropy} states that the cutoff for the adversary to be successful with high probability is still governed by the Shannon entropy (as intuitively expected).

In the regime where $g<H(\theta),$ we would like to study  the behavior of correct guessing. The next lemma relates the exponent of an exponentially large number of possible guesses  to the LDP rate function.
\begin{lemma}
If $g<H(\theta),$ then
\begin{equation}
\lim_{n \to \infty}\frac{1}{n} \log \frac{1}{\mathbb{P}_\theta[G_\theta(X^n) \leq e^{gn}]} = \Lambda^*_\theta(g).
\label{eq:metric2}
\end{equation}
\end{lemma}
Hence, $\mathbb{P}_\theta[G_\theta(X^n) \leq e^{gn}] \approx e^{-n \Lambda^*_\theta(g) }$, and a larger $\Lambda^*_\theta(g)$ directly implies a more secure source against a brute-force attacker who is subject to a guesswork budget $g$ for a fixed $n$. We use the above rate function as the metric for comparing two string-sources given a total guesswork budget, naturally defined as $g \times n$.

Using the notion of the tilt, we can represent the rate function $\Lambda_\theta^*(g)$ as a parametric function of $\alpha$ for a family of tilted distributions.
The rate function, $\Lambda_\theta^{*}(g)$, associated with $\theta \in \Theta_{|\mc{X}|} $ can be directly computed as~\cite{Beirami-IT-Guesswork}:
\begin{eqnarray}
\Lambda^{*}_\theta(g) = D\left( \tau(\theta,{\alpha(g)}) \Vert \theta \right),
\end{eqnarray}
for $\alpha(g) = \arg_{\alpha\in \mathbb{R}^+} \left\{ H(\tau(\theta,\alpha)) = g \right\}$. This characterization plays a central role in our derivations.

Recall that we adjust the string lengths in order to make sure that the secret string chosen by the user is subject to a given total entropy budget. 
As the idea of the total guesswork budget is that the adversary can make a fixed number of queries regardless of the source from which the user is choosing the password, we compare the sources in terms of the probability of success subject to an adjusted guesswork budget per source character (see~\eqref{eq:metric2}). To keep the total guessing budget of the adversary the same, i.e., $e^{n_1 g_1} = e^{n_2 g_2},$  we must adjust the guesswork budget per source character as follows:
\begin{equation}
    g_2 = \eta g_1.
\label{eq:ratio2}
\end{equation}
In light of~\eqref{eq:ratio2}, we compare $\Lambda^*_{\theta_1}(g_1)$ with $\frac{1}{\eta} \Lambda^*_{\theta_2}(g_2) = \frac{1}{\eta} \Lambda^*_{\theta_2}(\eta g_1)$ for sources with parameter vectors $\theta_1$ and $\theta_2$.

We are now ready to  provide our results on the adversary's probability of success.
\begin{theorem}
Let $\theta_1 \in \Theta_{|\mc{X}|}$. For any $\theta_2 \in \underline{\Gamma}^+_{\theta_1}$, 
\begin{equation}
 \Lambda_{\theta_1}^*\left(g_1\right) > \frac{1}{\eta} \Lambda_{\theta_2}^*\left(g_2\right), \quad \forall g_1<H(\theta_1),
\end{equation} 
if and only if $\theta_1$ satisfies the SEC (see Definition~\ref{def:information-skewness}). 
\label{thm:lambda}
\end{theorem}
We remark that the same SEC appears to be the crucial quantity for the statement of Theorem~\ref{thm:lambda} to hold. 
This theorem implies that when the adversary is subject to a guesswork budget $g_1$ (i.e., he can only submit $e^{n_1g_1}$ queries to identify a secret string of length $n$) for some $g_1\in(0, H(\theta_1))$, then the chances of correctly identifying the random string produced by a ``more uniform'' high per-character entropy member of the tilted family is exponentially smaller than that of the less uniform low per-character entropy source belonging to the same tilted family so long as the source satisfies the SEC when the user is subject to the same total entropy budget and the adversary is subject to the same total guesswork budget.
In particular, the uniform source is the most secure against such an adversary subject to a guesswork budget:

\begin{corollary}
Let $u_{|\mc{X}|}$ denote the uniform information source. Then, for any $\theta \in \Theta_{|\mc{X}|}$ and $g < \log |\mc{X}|$, we have
\begin{equation}
\log |\mc{X}| - g = \Lambda_{u_{|\mc{X}|}}^*\left(g\right) > \frac{1}{\eta} \Lambda_\theta^*(\eta g)  ,  
\end{equation} 
where $\eta= H(\theta)/ \log |\mc{X}|$.
\label{thm:lambda-uniform}
\end{corollary}
We remark that these security guarantees are against an adversary that is not powerful enough to be able to explore the entire typical set rendering his chances of success exponentially small. The ``more uniform'' sources provide an exponentially smaller chance to such an adversary to be successful.

We emphasize that the implications of Theorems~\ref{thm:avg_guesswork} and~\ref{thm:lambda} are in stark contrast to each other. On the one hand, more uniformity results in an exponential decrease in the  number of queries expected of an adversary to correctly identify a secret string when the user is subject to a total entropy budget (Theorem~\ref{thm:avg_guesswork}). On the other hand, more uniformity decreases the chances of an adversary in identifying the secret string when the adversary's power is limited by a total guesswork budget  as well (Theorem~\ref{thm:lambda}).

\section{Properties of the SEC}
\label{sec:skewness}
Noting that SEC introduced in Definition~\ref{def:information-skewness} is a new concept, we study this condition in more detail in this section.
Let us start with the binary memoryless sources. 
\begin{lemma}
Let $\theta \in \Theta_2$. Further,  let $\phi = \min\{\theta_1, \theta_2\}< \frac{1}{2}$. Then,
\begin{align}
H(\theta)& =  \phi \log \left(\frac{1}{\phi}\right) + (1-\phi) \log \left(\frac{1}{1-\phi}\right),\label{eq:H_binary}\\
V(\theta) & = \phi (1-\phi) \log^2 \left(\frac{1-\phi}{\phi}\right),\label{eq:V_binary}\\
S(\theta) & = \phi(1-\phi)(1-2\phi) \log^3 \left(\frac{1-\phi}{\phi}\right).\label{eq:S_binary}
\end{align} 
\end{lemma}
The next theorem is our main result for binary memoryless sources:
\begin{theorem}
Any $\theta \in \Theta_2$ satisfies the SEC.
\label{thm:skewness_binary}
\end{theorem}
While Theorem~\ref{thm:skewness_binary} shows that all binary memoryless sources satisfy the SEC, the same argument does not extend to larger alphabets. 
\begin{theorem}
For any $|\mc{X}|>2,$ there exists $\theta \in \Theta_{|\mc{X}|}$, such that $\theta$ does not satisfy the SEC.
\label{thm:large_alphabet}
\end{theorem}

\begin{figure}[tbp]  
  \centering
  \includegraphics[width = 0.7\textwidth]{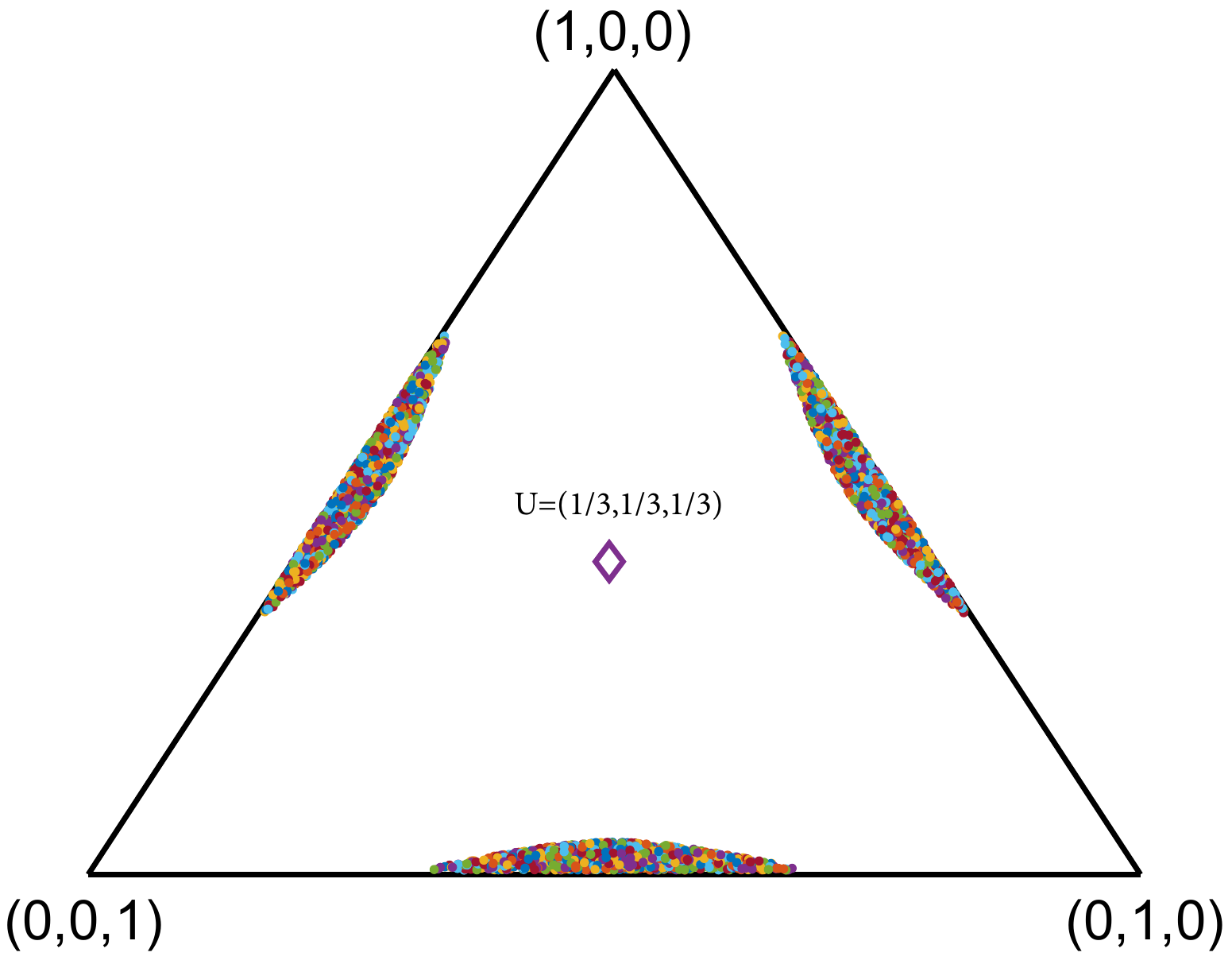}
  \caption{Depiction of the probability simplex for a ternary alphabet. The figure represents the set of distributions that do not satisfy the SEC.}
\label{fig:goodAndBadDistributions}
\end{figure}

Despite the negative result in Theorem~\ref{thm:large_alphabet}, we show that sources that are approximately uniform satisfy the SEC for any alphabet size. 
Here is the key result for such sources:
\begin{theorem}
Suppose that $\theta \in \Theta_{|\mc{X}|}$ is such that 
\begin{equation}
\left|\log \frac{1}{\theta_i} - H(\theta)\right| < 2, \quad \forall i \in [|\mc{X}|].
\label{eq:lem-skewness-uniform-condition1}
\end{equation}
Then $\theta$ satisfies the SEC.
\label{lem:sufficient-uniform1}
\end{theorem}
As a corollary, we state the condition more explicitly in terms of $\theta_i$'s.
\begin{corollary}
Suppose that $\theta \in \Theta_{|\mc{X}|}$ is such that 
\begin{equation}
\frac{e^{-1}}{|\mc{X}|}<\theta_i < \frac{e}{|\mc{X}|}, \quad \forall i \in [|\mc{X}|].
\label{eq:thm-skewness-uniform-condition}
\end{equation}
Then, $\theta$ satisfies the SEC. 
\label{thm:sufficient-uniform}
\end{corollary}

Figure~\ref{fig:goodAndBadDistributions} depicts the set of ternary distributions that do not satisfy the SEC. As can be seen, source close to uniform satisfy the SEC while sources that are close to uniform on a two-dimensional alphabet while almost missing the third character in the alphabet do not satisfy the SEC.

\section{Numerical Experiments}

\begin{figure}[t]  
  \centering
  \includegraphics[width = 0.7\textwidth]{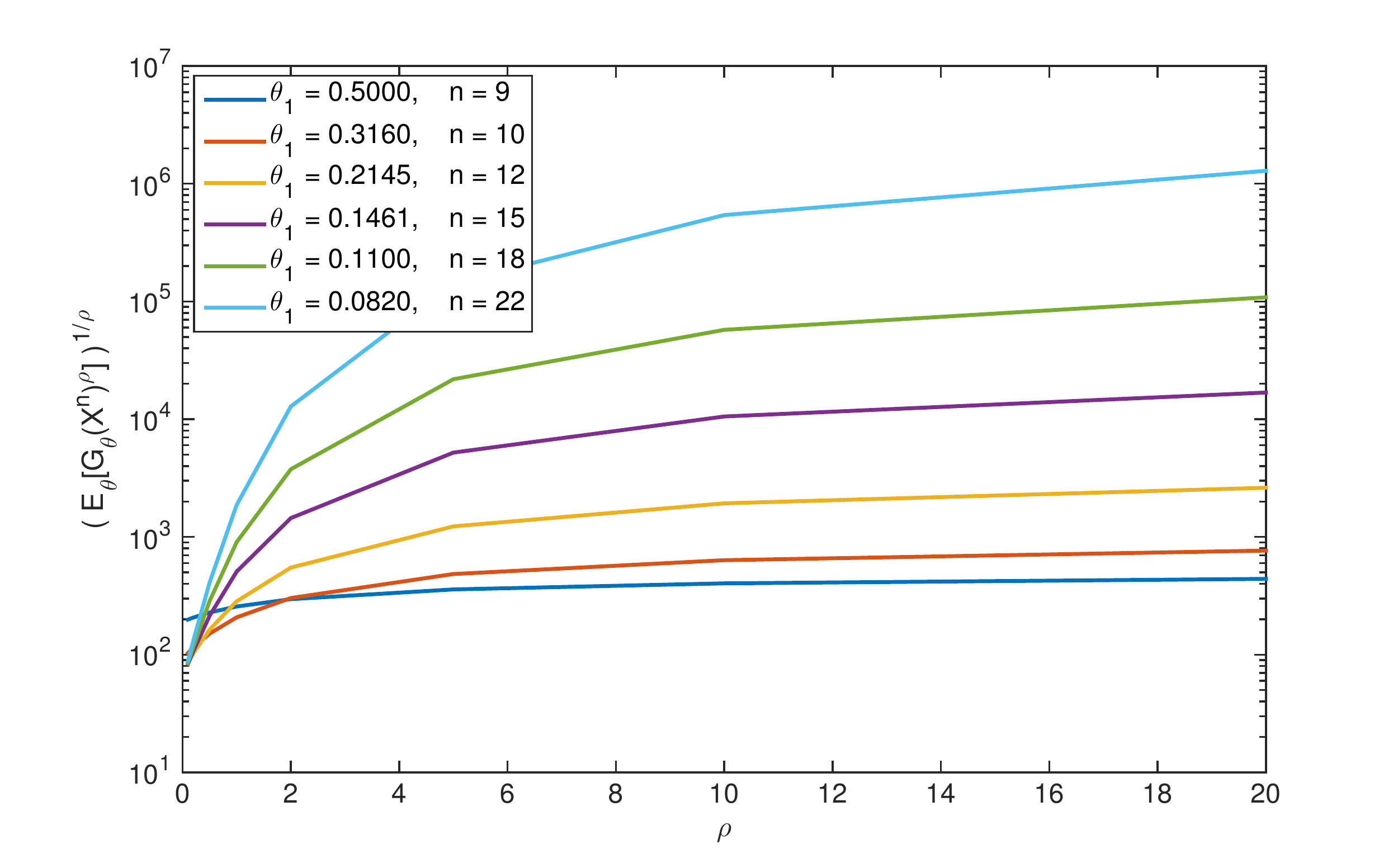}
  \caption{The positive moments of guesswork for sources subject to the same total entropy budget in Table~\ref{tab:sources}.}
\label{fig:theorem1}
\vspace{-.05in}
\end{figure}

\begin{figure}[t]  
  \centering
  \includegraphics[width = 0.7\textwidth]{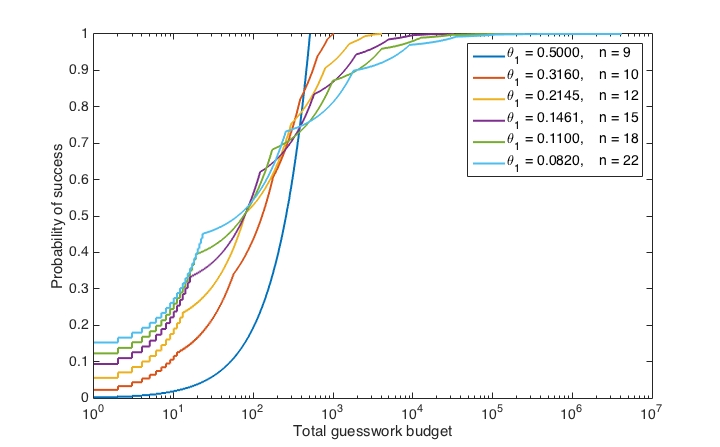}
  \caption{The probability of success as a function of the total guesswork budget for binary sources of Table~\ref{tab:sources} subject to the same total entropy budget.}
\label{fig:theorem2}
\vspace{-.05in}
\end{figure}

In this section, we provide some numerical experiments. We compare several binary sources, where $\theta = (\theta_1, \theta_2)$ is the source parameter vector.  
The parameter vectors used for the experiments are listed in Table~\ref{tab:sources}. The length and the parameter vector are chosen such that $n H(\theta) = 9 \log2$ nats for all of the pairs.
Although the theorems proved in this paper are of asymptotic nature, we have chosen to run experiments on finite-length sequences instead to emphasize the applicability of the results even in very short lengths. 
As can be seen in Fig.~\ref{fig:theorem1}, as the entropy rate of the source decreases, the moments of guesswork increase exponentially subject to the same entropy budget. On the other hand, as shown in Fig.~\ref{fig:theorem2}, as the entropy rate of the source decreases, the chances of an adversary subject to a fixed total guesswork budget increases, which is consistent with our intuition. 

\begin{table}[th]
\centering
\begin{tabular}{ |c|c| } 
 \hline
$ \theta_1$ & n  \\ 
 \hline
 0.5000 & 9 \\ 
0.3160 & 10  \\ 
0.2145 & 12  \\ 
0.1461& 15  \\ 
 0.1100 & 18  \\ 
 0.0820 & 22  \\ 
 \hline
\end{tabular}
\caption{The list of source parameters and sequence lengths of binary sources used in the experiments.}
\label{tab:sources}
\end{table}

\section{Conclusion}
In this paper, we studied guesswork  subject to a total entropy budget. We showed that the conclusions about security deduced from the analysis of the average guesswork could be counter-intuitive in that they suggest that the uniform source is not the strongest source against brute-force attacks. To remedy the problem, we introduced the concept of total guesswork budget, and showed that if the adversary is subject to a total guesswork budget, the uniform source provides the strongest security guarantees against the brute-force attacker, which is consistent with our intuition.

\appendix[Proofs]
\label{app:app}


\begin{IEEEproof}[Proof of Theorem~\ref{thm:avg_guesswork}]
This is equivalent to showing that for all $\rho>0$,
\begin{equation}
\frac{H_{1/(1+\rho)} (\theta_2)}{H(\theta_2)} > \frac{H_{1/(1+\rho)} (\theta_1)}{H(\theta_1)}
\end{equation}
for all $\theta_2 \in \underline{\Gamma}^+_\theta$. Let $\beta := 1/(1+\rho)$, and hence $\beta<1$.
The statement above is in turn equivalent to showing:
\begin{equation}
\frac{\partial}{\partial \alpha}\left[ \frac{H_{\beta} (\tau(\theta_1, \alpha))}{H(\tau(\theta_1, \alpha))}\right]_{\alpha = 1} > 0, \quad \forall \beta<1.
\label{eq:eq2}
\end{equation}
It is straightforward to show that~\eqref{eq:eq2} is equivalent to 
\begin{equation}
\frac{\frac{\partial }{\partial \alpha}\left[ H_{\beta} (\tau(\theta_1, \alpha))\right]_{\alpha = 1} }{{H_{\beta}(\theta_1)}}> \frac{\frac{\partial }{\partial \alpha}\left[ H (\tau(\theta_1, \alpha))\right]_{\alpha = 1} }{{H(\theta_1)}}, \quad \forall \beta<1.
\label{eq:eq3}
\end{equation}
Finally, we prove the following statement that is equivalent to~\eqref{eq:eq3}:
\begin{equation}
\frac{\partial}{\partial \beta}\left[\frac{\frac{\partial}{\partial \alpha}\left[ H_{\beta} (\tau(\theta_1, \alpha))\right]_{\alpha = 1} }{{H_{\beta}(\theta_1)}}\right]_{\beta=1}<0.
\label{eq:eq4}
\end{equation}
This is equivalent to showing:
\begin{align}
&\frac{\partial^2}{\partial \alpha \partial \beta}\left[H_{\beta} (\tau(\theta_1, \alpha)) \right]_{\alpha = \beta=1} \quad H (\theta_1) \nonumber\\
<
&\frac{\partial}{\partial \beta} \left[H_{\beta} (\theta_1) \right]_{\beta = 1} \quad
\frac{\partial}{\partial \alpha }\left[H (\tau(\theta_1, \alpha)) \right]_{\alpha =1}.
\label{eq:eq5}
\end{align}
The above statement is shown to hold if and only if $\theta_1$ satisfies the SEC (Definition~\ref{def:information-skewness}) invoking Lemmas~\ref{lem:H-da},~\ref{lem:H-db}, and~\ref{lem:H-dd}, which completes the proof of the theorem.
\end{IEEEproof}
\begin{lemma}
For all $\theta \in \Theta_{|\mc{X}|}$, we have
\begin{equation}
\frac{\partial}{\partial \alpha }\left[H (\tau(\theta, \alpha)) \right]_{\alpha = 1} = - V(\theta).
\end{equation}
\label{lem:H-da}
\end{lemma}
See~\cite{Beirami-IT-Guesswork} for the proof.
\begin{lemma}
For all $\theta \in \Theta_{|\mc{X}|}$, we have
\begin{equation}
\frac{\partial}{ \partial \beta}\left[H_{\beta} (\theta) \right]_{\beta =1 } = -\frac{1}{2} V(\theta).
\end{equation}
\label{lem:H-db}
\end{lemma}
See~\cite{Beirami-IT-Guesswork} for the proof.
\begin{lemma}
For all $\theta \in \Theta_{|\mc{X}|}$, we have
\begin{equation}
\frac{\partial^2}{\partial \alpha \partial \beta}\left[H_{\beta} (\tau(\theta, \alpha)) \right]_{\alpha = \beta =1} = - V(\theta) +\frac{1}{2} S(\theta).
\end{equation}
\label{lem:H-dd}
\end{lemma}
\begin{IEEEproof}
It is proved in~\cite{Beirami-IT-Guesswork} that
\begin{equation}
\frac{\partial}{\partial \alpha }\left[H_{\beta} (\tau(\theta, \alpha)) \right]_{\alpha = 1} = \frac{\beta}{1-\beta} \left( H(\theta) - H(\tau(\theta, \beta)||\theta)  \right).
\end{equation}
Hence, we differentiate with respect to $\beta$ to get:
\begin{align*}
\frac{\partial^2}{\partial \alpha \partial \beta }\left[H_{\beta} (\tau(\theta, \alpha)) \right]_{\alpha = 1}& = \frac{1}{(1-\beta)^2} \left( H(\theta) - H(\tau(\theta, \beta)||\theta)  \right) \\
& +\frac{\beta}{1-\beta}V(\tau(\theta, \beta)||\theta) .
\end{align*}
Next, we take the limit as $\beta \to 1$, and by applying  L'Hospital's rule we arrive at:
\begin{equation}
\frac{\partial^2}{\partial \alpha \partial \beta}\left[H_{\beta} (\tau(\theta, \alpha)) \right]_{\alpha = \beta =1} = - V(\theta) -\frac{1}{2} \frac{\partial}{\partial\beta} [V(\tau(\theta, \beta)||\theta)]_{\beta=1}.
\end{equation}
Finally, the proof is completed by invoking Lemma~\ref{lem:dV}.
\end{IEEEproof}

\begin{lemma}
For any $\theta \in \Theta_{|\mc{X}|}$, 
$$
\frac{\partial}{\partial\alpha} [V(\tau(\theta, \alpha)||\theta)]_{\alpha=1} = - S(\theta),
$$
where $S(\theta)$ is defined in~\eqref{eq:skewness}.
\label{lem:dV}
\end{lemma}
\begin{IEEEproof}
By definition
\begin{align}
&\hspace{-.1in}\left.\frac{\partial }{\partial \alpha}V(\tau( \theta, \alpha)||\theta)\right|_{\alpha = 1} \nonumber\vspace{0.1in}\\
&= \sum_{i \in [|\mc{X}|]}\left.\frac{\partial}{\partial\alpha}  \tau_i( \theta, \alpha) \right|_{\alpha=1} \left(H(\tau(\theta,\alpha)||\theta) - \log \frac{1}{\theta_i}  \right)^2\nonumber\\
&= \sum_{i \in [|\mc{X}|]}\theta_i \left(H(\tau(\theta,\alpha)||\theta) - \log \frac{1}{\theta_i}  \right)^3\label{eq:lllz}\\
& = -S(\theta),
\end{align}
where~\eqref{eq:lllz} follows by invoking Lemma 8 of~\cite{Beirami-IT-Guesswork}.
\end{IEEEproof}

\begin{IEEEproof}[Proof of Theorem~\ref{thm:lambda}]
Let us recall that $\theta_2 = \tau(\theta_1,\alpha) $ for some $\alpha>1$. We can find $t_1$ and $t_2$ in the domain of each rate function such that the derivatives of the rate function are both equal to a constant $\rho>-1$. It follows from~\cite{arikan1996inequality} that:
\begin{align}
t_1 =  \arg_{t}  \left\{ \frac{\partial}{\partial t} \Lambda_{\theta_1}^*(t) = \rho \right\}  & \Rightarrow  t_1 =  H(\tau(\theta_1, \beta)), \nonumber \\
t_2  =  \arg_{t}  \left\{ \frac{1}{\eta} \frac{\partial }{\partial t} \Lambda_{\theta_2}^*\left(\eta t\right)  = \rho \right\} & \Rightarrow  t_2 = \frac{1}{\eta} H(\tau(\theta_2, \beta)) \label{eq:t2},
\end{align}
where $\beta = 1/(1+\rho)$. We focus on $\rho<0$, and hence $\beta \in (1,\infty)$. Note that $\beta = 1$, (equivalently $\rho=0$) corresponds to the coinciding zeros of both rate functions. Once again recalling that the rate functions are convex, proving $(1/\eta) \Lambda_{\theta_2}^*(\eta t) > \Lambda_{\theta_1}^*(t)$ is equivalent to showing that $t_2 < t_1$ (as defined in~\eqref{eq:t2}) for all $\beta >1$. This is in turn equivalent to showing:
\begin{eqnarray}
\frac{H(\tau(\theta_2,  \beta))}{H(\theta_2)}  < \frac{H(\tau(\theta_1,{\beta }))}{H(\theta_1)} , \hspace{0.5cm} \forall \alpha,\beta > 1.
\end{eqnarray}
This is equivalent to:
\begin{equation}
\frac{\partial}{\partial \alpha} \left[\frac{H(\tau(\theta_1, \alpha \beta))}{H(\tau(\theta_1,\alpha))} \right]_{\alpha=1} <0, \quad \forall \beta>1.
\label{eq:eq22}
\end{equation}
It is straightforward to show that~\eqref{eq:eq22} is equivalent to 
\begin{equation}
\frac{\frac{\partial }{\partial \alpha}\left[ H (\tau(\theta_1, \alpha\beta))\right]_{\alpha = 1} }{{H(\tau(\theta_1,\beta))}}> \frac{\frac{\partial }{\partial \alpha}\left[ H (\tau(\theta_1, \alpha))\right]_{\alpha = 1} }{{H(\theta_1)}}, \quad \forall \beta>1.
\label{eq:eq23}
\end{equation}
Finally, we prove the following statement that is equivalent to~\eqref{eq:eq23}:
\begin{equation}
\frac{\partial}{\partial \beta}\left[\frac{\frac{\partial}{\partial \alpha}\left[ H (\tau(\theta_1, \alpha\beta))\right]_{\alpha = 1} }{{H(\tau(\theta_1,\beta))}}\right]_{\beta=1}<0.
\label{eq:eq24}
\end{equation}
This is equivalent to showing:
\begin{align}
&\frac{\partial^2}{\partial \alpha \partial \beta}\left[H (\tau(\theta_1, \alpha\beta)) \right]_{\alpha = \beta=1} \quad H (\theta_1)\nonumber \\
<
&\frac{\partial}{\partial \beta} \left[H (\tau(\theta_1,\beta)) \right]_{\beta = 1} \quad
\frac{\partial}{\partial \alpha }\left[H (\tau(\theta_1, \alpha)) \right]_{\alpha =1}.
\label{eq:eq25}
\end{align}
The above statement is shown to hold if and only if $\theta_1$ satisfies the SEC (Definition~\ref{def:information-skewness}) invoking Lemmas~\ref{lem:H-da} and~\ref{lem:H-dab}, which completes the proof of the theorem.
\end{IEEEproof}

\begin{lemma}
For all $\theta \in \Theta_{|\mc{X}|}$, we have
\begin{equation}
\frac{\partial^2}{\partial \alpha \partial \beta}\left[H (\tau(\theta, \alpha\beta)) \right]_{\alpha = \beta =1} = - 2V(\theta) + S(\theta).
\end{equation}
\label{lem:H-dab}
\end{lemma}
\begin{IEEEproof}
Noting that $\tau(\theta, \alpha\beta) = \tau(\tau(\theta, \beta),\alpha)$ and invoking Lemma~\ref{lem:H-da}, we have
\begin{align}
\frac{\partial}{\partial \alpha }\left[H (\tau(\theta, \alpha\beta)) \right]_{\alpha = 1} &= - V(\tau(\theta,\beta))\\
& = - \beta^2 V(\tau(\theta,\beta)||\theta), \label{eq:zz}
\end{align}
where~\eqref{eq:zz} follows from Lemma 5 of~\cite{Beirami-IT-Guesswork}.
Hence, by differentiating the above with respect to $\beta$ at $\beta=1$ and invoking Lemma~\ref{lem:dV}, we arrive at the claim.
\end{IEEEproof}

\begin{IEEEproof}[Proof of Theorem~\ref{thm:skewness_binary}]
The theorem is proved by invoking Lemmas~\ref{lem:SH_bound} and~\ref{lem:HV_bound}, as follows:
\begin{align}
H(\theta) S(\theta) &<  V^2(\theta) + \phi^2(1-\phi) (1-2\phi) \log^3\left(\frac{1-\phi}{\phi}\right)\\
&< V^2(\theta) + V(\theta) H(\theta)\\
& <V^2(\theta) + 2V(\theta) H(\theta),
\end{align}
and hence $\theta$ satisfies the SEC.
\end{IEEEproof}

\begin{lemma}
For any $\theta \in \Theta_2$, we have
\begin{equation}
H(\theta) S(\theta) <  V^2(\theta) + \phi^2(1-\phi) (1-2\phi) \log^3\left(\frac{1-\phi}{\phi}\right),
\end{equation}
where $\phi := \min\{\theta_1, \theta_2\}$.
\label{lem:SH_bound}
\end{lemma}
\begin{IEEEproof}
Let $\phi = \min\{\theta_1, \theta_2\}$.
First note that by Lemma~\ref{lem:xlogx}, we have
\begin{equation}
H(\theta)<\phi\log \frac{1}{\phi}  + \phi.
\end{equation}
Hence,
\begin{align}
H(\theta ) S(\theta) &<\phi^2(1-\phi)(1-2\phi) \log^3 \left(\frac{1-\phi}{\phi}\right)\nonumber\\
&+ \phi^2(1-\phi)(1-2\phi) \log^3 \left(\frac{1-\phi}{\theta}\right) \log\left(\frac{1}{\phi}\right)\\
&<\phi^2(1-\phi)(1-2\phi) \log^3 \left(\frac{1-\phi}{\phi}\right)\nonumber\\
&+ \phi^2(1-\phi)^2 \log^4 \left(\frac{1-\phi}{\phi}\right), \label{eq:H_bound}
\end{align}
where \eqref{eq:H_bound} follows from Lemma~\ref{lem:xlogx2}, completing the proof.
\end{IEEEproof}

\begin{lemma}
For any $\theta \in \Theta_2$, we have 
\begin{equation}
H(\theta)V(\theta)>\phi^2(1-\phi) (1-2\phi) \log^3\left(\frac{1-\phi}{\phi}\right), 
\end{equation}
where $\phi: = \min\{\theta_1, \theta_2\}.$
\label{lem:HV_bound}
\end{lemma}
\begin{IEEEproof}
For $\phi = \min\{\theta_1, \theta_2\},$ note that 
\begin{equation}
H(\theta) > \phi \log \frac{1}{\phi},
\end{equation}
and hence
\begin{align}
H(\theta) V(\theta) &> \phi^2 (1-\phi) \log^2\left(\frac{1-\phi}{\phi}\right) \log \left(\frac{1}{\phi}\right)\\
& >\phi^2(1-\phi) (1-2\phi) \log^3\left(\frac{1-\phi}{\phi}\right),\label{eq:xlogx2app}
\end{align}
where~\eqref{eq:xlogx2app} follows from Lemma~\ref{lem:xlogx2}, completing the proof.
\end{IEEEproof}

\begin{lemma}
For any $0<x<1$, we have
\begin{equation}
(1-x) \log \frac{1}{1-x} <x.
\end{equation}
\label{lem:xlogx}
\end{lemma}
\begin{IEEEproof}
Note that as $x\to 0$ both sides are equal and the limit of their derivatives are equal as well, while the second derivative of the left hand side is equal to $- \frac{1}{1-x} <0$ completing the proof.
\end{IEEEproof}

\begin{lemma}
For any $0<x<\frac{1}{2}$, we have
\begin{equation}
(1-2x) \log \frac{1}{x} <(1-x) \log \frac{1-x}{x} .
\end{equation}
\label{lem:xlogx2}
\end{lemma}
\begin{IEEEproof}
The proof is similar to that of Lemma~\ref{lem:xlogx}.
\end{IEEEproof}

\begin{IEEEproof}[Proof of Theorem~\ref{thm:large_alphabet}]
We proceed with the proof by construction.
Let $\theta$ be such that 
\begin{equation}
\theta_i = \left\{
\begin{array}{ll}
(1-\epsilon)/(|\mc{X}|-1) & 1 \leq i \leq |\mc{X}| - 1\\
\epsilon & {i = |\mc{X}|}
\end{array}
\right..
\end{equation}
Then, invoking Lemma~\ref{lem:HVS_construct}, we can see that as $\epsilon \to 0$, for sufficiently small $\epsilon$ and $|\mc{X}|>2$, we have
\begin{align}
\frac{1}{2} \log (|\mc{X}|-1) < &H(\theta) <2 \log (|\mc{X}|-1),\\
\frac{1}{2}\epsilon \left(\log \frac{1}{\epsilon}\right)^2 <&V(\theta) <\epsilon \left(\log \frac{1}{\epsilon}\right)^2,\\
\frac{1}{2}\epsilon \left(\log \frac{1}{\epsilon}\right)^3 <&S(\theta) <\epsilon \left(\log \frac{1}{\epsilon}\right)^3.
\end{align}
Hence, 
\begin{align}
S(\theta)H(\theta)&> \frac{1}{4}\epsilon \left(\log \frac{1}{\epsilon}\right)^3  \log (|\mc{X}|-1) \\
& >\epsilon^2 \left(\log \frac{1}{\epsilon}\right)^4+ 4\epsilon \left(\log \frac{1}{\epsilon}\right)^2\log (|\mc{X}|-1) \label{eq:suff-small}\\
&> V^2(\theta) + 2H(\theta) V(\theta).
\end{align}
where~\eqref{eq:suff-small} holds for sufficiently small $\epsilon$ as long as $|\mc{X}|>2$. Thus, $\theta$ does not satisfy the SEC, and the proof is complete.
\end{IEEEproof}

\begin{lemma}
\label{lem:HVS_construct}
Let $\theta \in \Theta_{|\mc{X}|}$ be such that 
\begin{equation}
\theta_i = \left\{
\begin{array}{ll}
(1-\epsilon)/(|\mc{X}|-1) & 1 \leq i \leq |\mc{X}| - 1\\
\epsilon & {i = |\mc{X}|}
\end{array}
\right..
\end{equation}
Then,
\begin{align}
H(\theta)& =  (1-\epsilon) \log (|\mc{X}|-1) + h(\epsilon),\label{eq:H_construct}\\
V(\theta) & = \epsilon (1-\epsilon) \left(\log\left( \frac{1-\epsilon}{\epsilon }\right)  - \log (|\mc{X}|-1)\right)^2,\label{eq:V_construct}\\
S(\theta) & = \epsilon(1-\epsilon)(1-2\epsilon) \left(\log\left( \frac{1-\epsilon}{\epsilon }\right)  - \log (|\mc{X}|-1) \right)^3,\label{eq:S_construct}
\end{align}
where $h(\epsilon)$ is the binary entropy function given by
\begin{equation}
h(\epsilon): = H(\epsilon, 1-\epsilon) = \epsilon \log \frac{1}{\epsilon} + (1-\epsilon) \log \frac{1}{1-\epsilon}.
\label{eq:binary_entropy}
\end{equation}
\end{lemma}
\begin{IEEEproof}
The calculation of $H(\theta)$ is straightforward by noting that this is a mixture of two uniform sources on alphabets of size $(|\mc{X}|-1)$ and $1$. To calculate $V(\theta)$, we have
\begin{align}
V(\theta) &=(1-\epsilon) \left( \log \frac{|\mc{X}|-1}{1-\epsilon}  - (1-\epsilon)\log (|\mc{X}|-1) -  h(\epsilon)\right)^2\nonumber \\
&+ \epsilon \left( \log \frac{1}{\epsilon} -  (1-\epsilon)\log (|\mc{X}|-1) -  h(\epsilon) \right)^2\\
 &=(1-\epsilon) \left( \epsilon \log (|\mc{X}|-1) + \epsilon \log \frac{\epsilon}{1-\epsilon} \right)^2\nonumber \\
&+ \epsilon \left( -(1-\epsilon)  \log (|\mc{X}|-1) +(1- \epsilon) \log \frac{1-\epsilon}{\epsilon} \right)^2\\
& = \epsilon (1-\epsilon) \left(  \log \frac{1-\epsilon}{\epsilon} - \log (|\mc{X}|-1) \right)^2.
\end{align}
Finally, to calculate $S(\theta)$, similarly to the calculations for $V(\theta)$, we get
\begin{align}
S(\theta)  &=(1-\epsilon) \left( \epsilon \log (|\mc{X}|-1) + \epsilon \log \frac{\epsilon}{1-\epsilon} \right)^3\nonumber \\
&+ \epsilon \left( -(1-\epsilon)  \log (|\mc{X}|-1) +(1- \epsilon) \log \frac{1-\epsilon}{\epsilon} \right)^3\\
& = \epsilon (1-\epsilon)(1-2\epsilon) \left(  \log \frac{1-\epsilon}{\epsilon} - \log (|\mc{X}|-1) \right)^3,
\end{align}
establishing the claim.
\end{IEEEproof}

\begin{IEEEproof}[Proof of Theorem~\ref{lem:sufficient-uniform1}]
Let $X$ be drawn from $\theta$. Further, let  $$Y = \log \frac{1}{P(X)} - H(X).$$ Hence, by definition, $E[Y^3] = S(\theta)$ and $E[Y^2] = V(\theta)$.
Then, the condition in~\eqref{eq:lem-skewness-uniform-condition1} would ensure that $Y \in [-2,2]$. Noting that the uniform distribution is excluded in $\Theta_{|\mc{X}|}$, and hence the varentropy is nonzero, we apply Lemma~\ref{lem:skew} (with $a =2$) to obtain that
$$
S(\theta)<2V(\theta).
$$
This is a sufficient condition for the SEC to hold, completing the proof.
\end{IEEEproof}
\begin{lemma}
Let $Y$ be a random variable supported on $[-a,a]$ for some $a >0$ Further, let $E[Y] = 0$ and $E[ Y^2]> 0$. Then, 
\begin{equation}
\frac{E [Y^3]}{E [Y^2]}\leq a.
\end{equation}
\label{lem:skew}
\end{lemma}
\begin{IEEEproof}
It is straightforward to show that $\frac{E [Y^3]}{E [Y^2]}$ is maximized if 
$$
p_y(y) = \left\{ 
\begin{array}{ll}
\rho/ 2,& y = -a\\
1-\rho,& y = 0\\
\rho/2,& y = a
\end{array}
\right.,
$$
for some $\rho>0$, which in turn leads to $\frac{E [Y^3]}{E [Y^2]} = a.$
\end{IEEEproof}

\begin{IEEEproof}[Proof of Corollary~\ref{thm:sufficient-uniform}]
First we show that the condition in~\eqref{eq:thm-skewness-uniform-condition} leads to the condition in~\eqref{eq:lem-skewness-uniform-condition1}, which follows from the following set of inequalities:
\begin{align}
\max_{i \in [|\mc{X}|]}\left|\log \frac{1}{\theta_i} - H(\theta)\right| &\leq \max_{i \in [|\mc{X}|]} \left|\log \frac{1}{\theta_i} - \log |\mc{X}|\right| \nonumber\\
& + \left|\log|\mc{X}|- H(\theta)\right| \label{eq:eq1}\\
& \leq 2\max_{i \in [|\mc{X}|]} \left|\log \frac{1}{\theta_i} - \log |\mc{X}|\right|\\
& = 2,\label{eq:eq2}
\end{align}
where~\eqref{eq:eq1} follows Jensen's inequality and the convexity of the $|\cdot|$ operator, and~\eqref{eq:eq2} is a direct result of~\eqref{eq:thm-skewness-uniform-condition}. Hence, the claim of Lemma~\ref{lem:sufficient-uniform1} holds, which results in the claim of the theorem.
\end{IEEEproof}

\bibliographystyle{IEEEtran}
\bibliography{IEEEabrv,references}

\end{document}